\documentclass[11pt,a4paper]{article}

\usepackage[utf8]{inputenc}
\usepackage[T1]{fontenc}
\usepackage{lmodern,amsmath,amssymb,graphicx,authblk,booktabs,subcaption}
\usepackage[margin=1in]{geometry}
\usepackage{microtype}
\usepackage{hyperref}
\hypersetup{colorlinks=true,linkcolor=blue,citecolor=blue,urlcolor=blue}

\usepackage{cite}

\newcommand{\eighttev}[1]{8tev_bfklVsPythiaVsHerwig/#1}
\newcommand{\thirteentev}[1]{13tev_bfklVsPythiaVsHerwig/#1}

\title{\vspace{-0.5em}Order statistics for multijet events\vspace{0.25em}}

\author[1]{G.~Chachamis}
\author[2,3]{A.~Sabio Vera}
\author[4,5]{D.~Vaccaro}

\affil[1]{Departamento de Ciencias, Universidad Pública de Navarra, Campus de Arrosadía, 31006 Pamplona, Spain}
\affil[2]{Instituto de Física Teórica UAM/CSIC, Nicolás Cabrera 15, E-28049 Madrid, Spain}
\affil[3]{Theoretical Physics Department, Universidad Autónoma de Madrid, E-28049 Madrid, Spain}
\affil[4]{Laboratório de Instrumentação e Física Experimental de Partículas (LIP), Av.\ Prof.\ Gama Pinto, 2, P-1649-003 Lisboa, Portugal}
\affil[5]{Departamento de Física, Instituto Superior Técnico, Universidade de Lisboa,
Av. Rovisco Pais 1, 1049-001 Lisbon, Portugal}

\date{}

\begin{document}
\maketitle

\begin{abstract}
We show that rank--ordered jet rapidity distributions---a direct application of order statistics---provide a simple yet powerful probe of
high--energy (small-$x$) QCD dynamics at the LHC. In inclusive dijet topologies at $\sqrt{s}=8$ and $13$~TeV, with realistic jet 
selections, we compare a BFKL--based Monte Carlo (\textsc{BFKLex}) 
to two general--purpose event generators based on collinear factorization 
and DGLAP parton showers, \textsc{PYTHIA8} ($p_T$--ordered) 
and \textsc{HERWIG7} (angular--ordered).
Even when two underlying dynamics happen to give similar \emph{inclusive} 
jet rapidity distributions, such observables are too coarse to 
discriminate their underlying rapidity point processes, whereas 
the \emph{rank--ordered} distributions remain sensitive to the 
differences in how rapidity space is filled.
For fixed multiplicity ($N{=}3$) and for the second--most--forward/backward jets across multiplicities, \textsc{BFKLex} 
populates the rapidity interval more democratically, whereas the general--purpose event generators 
exhibit comparatively stronger edge enhancement for $N{=}3$ and narrower, more centrally concentrated distributions for the second--most ranks.
These shape differences are stable under variations of jet radius, proton PDFs, and MPI/hadronization settings, 
and persist when requiring large rapidity separation between the outer jets. Rank--ordered rapidities thus 
compress genuinely \emph{exclusive} information about the multi--jet final state into one--dimensional, 
normalized histograms that are directly measurable with existing dijet and Mueller--Navelet selections 
and provide a new handle on high--energy radiation patterns.
\end{abstract}

\vspace{0.5em}

\section{Introduction}

Understanding the emergent properties of QCD in the high--energy (small--$x$) regime 
is an important goal of the LHC program. 
When the center--of--mass energy of the hard scattering is much larger than any other hard scale, 
the perturbative series develops enhanced terms $\alpha_s^n \ln^n s$ which are 
resummed in the Balitsky--Fadin--Kuraev--Lipatov (BFKL) framework~\cite{Lipatov:1985uk,Balitsky:1978ic,Kuraev:1977fs,Kuraev:1976ge,Lipatov:1976zz,Fadin:1975cb}. 
Mueller--Navelet (MN) dijets---two high-$p_T$ jets produced with a large 
rapidity separation and additional radiation treated inclusively---have long 
been proposed~\cite{Mueller:1986ey} as a promising laboratory to look for 
BFKL dynamics. More generally, inclusive dijet topologies with extra radiation 
offer a broad and experimentally accessible setting in which to probe 
high--energy QCD.

Over the last decade, several LHC measurements of these 
quantities have been found to be consistent both with NLL 
BFKL calculations and with fixed--order QCD supplemented by DGLAP--based parton showers. 
In hindsight this is not surprising: most of the standard Mueller--Navelet observables are either fully 
inclusive, or depend only on low--dimensional 
projections (such as single--jet distributions or a single azimuthal angle) 
of a complicated multi--jet final state. Very different underlying point processes 
in rapidity can share the same inclusive jet density $dN_{\text{jets}}/dy$ and 
similar azimuthal moments. What is missing, therefore, is an observable that is 
still as simple as a one--dimensional histogram, but that is built to 
retain genuinely \emph{exclusive} information about how jets populate the inter--tag rapidity interval.

In this Letter we propose exactly such a class of observables. 
We revisit the multiperipheral picture of high--energy scattering, 
in which produced particles are strongly ordered in rapidity 
and only weakly correlated in transverse momentum, and 
recast it in the language of \emph{order statistics}. Concretely, 
we treat the set of jet rapidities in an event as a finite sample 
drawn from an effective parent density in rapidity space, and 
study the distributions of the ordered rapidities (most backward, second--most backward, \dots, most forward). 
This order--statistics view makes precise a simple but powerful statement: 
if two dynamics induce different ``parent'' rapidity patterns, they \emph{must} 
differ in at least some of the rank--ordered distributions, even when their inclusive $dN_{\text{jets}}/dy$ happen to look similar.

Our main result is that rank--ordered rapidity distributions 
expose robust shape differences between a BFKL--based generator 
and standard DGLAP showers~\cite{Gribov:1972ri,Gribov:1972rt,Lipatov:1974qm,Altarelli:1977zs,Dokshitzer:1977sg} 
under realistic LHC conditions. These differences appear already at $p_{T,\min}\simeq 20$~GeV, 
survive modest variations of jet radius and PDFs, and persist when 
restricting to events with sizable rapidity spans. Because 
the observables are normalized one--dimensional histograms, they are simple to measure 
with existing MN selections and lend themselves to straightforward comparison between data and theory.

\section{Order statistics for jet rapidities}

Consider an event with $N$ reconstructed jets in a rapidity 
interval $I=[y_{\min},y_{\max}]$. Let $Y_1,\dots,Y_N$ denote their rapidities, 
and define the ordered rapidities
\begin{equation}
  Y_{(1)} < Y_{(2)} < \dots < Y_{(N)}\,,
\end{equation}
so that $Y_{(1)}$ and $Y_{(N)}$ are the most backward and most forward jets, respectively.

The usual inclusive jet rapidity density $dN_{\text{jets}}/dy$ is a 
\emph{marginal}: it counts jets in rapidity bins, summed over all 
multiplicities and over all multi--jet configurations. 
As such, it does not retain information about how jets are distributed 
\emph{within} a given event. Rank--ordered rapidities $Y_{(1)},\dots,Y_{(N)}$,
by contrast, probe the joint rapidity structure of the event in a way that
is still summarized by one--dimensional histograms.

A simple toy example makes this loss of information explicit. 
Consider events with exactly two jets in a symmetric rapidity window. 
In ``Model~A'', each jet is placed independently and uniformly in the window. 
In ``Model~B'', the jets are always produced in perfectly symmetric pairs at 
$y=\pm U$, with $U$ uniform in the forward half of the detector. 
By construction, both models have the same single--jet distribution 
$dN_{\text{jets}}/dy$: every rapidity bin is equally populated on average. 
Yet their ordered configurations differ dramatically: the most forward and 
most backward jets in Model~B tend to lie close to the edges and span a large 
rapidity interval, whereas in Model~A they are typically closer together. 
The two models are therefore indistinguishable at the level of 
$dN_{\text{jets}}/dy$ but are trivially distinguished by the distributions of 
$Y_{(1)}$, $Y_{(2)}$ and their difference $Y_{(2)}-Y_{(1)}$.

Rank--ordered observables such as $Y_{(1)},Y_{(2)},\dots,Y_{(N)}$ thus 
provide simple one--dimensional probes of the \emph{joint} rapidity 
structure of the event. They compress multi--particle exclusive 
information into a handful of normalized histograms that are as 
easy to measure and compare as $dN_{\text{jets}}/dy$, but remain 
sensitive to correlations and production patterns that the inclusive 
density averages away. This can be formalized in an idealized i.i.d.\ limit.

Suppose the single--jet rapidities are approximately i.i.d.\ draws from a 
parent density $f(y)$ on $I$ with cumulative distribution $F(y)$. 
The density of the $\ell$th order statistic $Y_{(\ell)}$ is the textbook result
\begin{equation}
  f_{(\ell)}(y)
  = \frac{N!}{(\ell-1)!(N-\ell)!}\,[F(y)]^{\ell-1}\,[1-F(y)]^{N-\ell}\,f(y)\,,
  \qquad \ell=1,\dots,N.
  \label{eq:order}
\end{equation}
Once the parent law $f(y)$ is fixed, the \emph{shape} of each ranked rapidity 
spectrum is fully determined. For a uniform parent on $I$ (a ``rapidity plateau''), 
the ordered rapidities map to Beta distributions in an affine variable $x\in[0,1]$, 
with outer ranks piling up toward the edges and inner ranks remaining broad. 
Any departure from a plateau---for example, an edge--enhanced or centrally peaked parent---induces characteristic, 
rank--dependent distortions of these shapes.

In an LHC analysis, jets are only reconstructed inside a finite window $\mathcal{W}=[-4.7,4.7]$. 
If the produced parent density on the real line is $f_{\rm prod}(y)$, 
the \emph{observable} parent within $\mathcal{W}$ is the truncated law
\begin{equation}
  f_{\rm obs}(y)
  = \frac{f_{\rm prod}(y)}{\displaystyle\int_{y_{\min}}^{y_{\max}} f_{\rm prod}(u)\,du}\,,
  \qquad y\in\mathcal{W},
\end{equation}
and Eq.~\eqref{eq:order} holds with $f\to f_{\rm obs}$ and $F\to F_{\rm obs}$ 
for events with exactly $N$ tagged jets in $\mathcal{W}$. 
A transverse--momentum threshold $p_{T,\min}$ affects $f_{\rm obs}$ only through 
the rapidity dependence of the pass probability $\Pr(p_T>p_{T,\min}\!\mid Y=y)$.

From this perspective, rank--ordered rapidity histograms are order--statistics 
projections of the effective parent law realized by a given dynamics. 
In a multiperipheral/BFKL scenario one expects a relatively broad, plateau--like parent, 
whereas DGLAP showers with coherence and recoil generically generate narrower parents.

\section{Monte Carlo setup and observables}

We study proton--proton collisions at $\sqrt{s}=8$ and $13$~TeV in inclusive dijet topologies. 
Jets are reconstructed with the anti-$k_T$ algorithm, as implemented in the \textsc{FastJet} 
package~\cite{Cacciari:2011ma}, with radius parameter $R=0.5$ at 8~TeV and $R=0.4$ at 13~TeV, 
and are required to lie in $|y|<4.7$ with $p_T>20$~GeV. All jets passing these cuts are ordered by rapidity.

We analyse two families of normalized histograms:
\begin{itemize}
  \item \emph{Fixed multiplicity $N=3$.} We consider events with 
  exactly three jets and measure the rapidity distributions of
  \[
    \text{Jet1} \equiv Y_{(1)},\quad
    \text{Jet2} \equiv Y_{(2)},\quad
    \text{Jet3} \equiv Y_{(3)}\,.
  \]
  \item \emph{Across multiplicities.} Over all multiplicities $N\ge 2$, we consider the \emph{most backward} (MB), \emph{most forward} (MF), and the \emph{second--most backward} (SMB) and \emph{second--most forward} (SMF) jets, defined as $Y_{(1)}$, $Y_{(N)}$, $Y_{(2)}$ and $Y_{(N-1)}$, respectively, whenever the corresponding rank exists.
\end{itemize}
In all cases the histograms are normalized to unit area, so only shapes matter.

For the dynamics, we compare three generators:
\begin{enumerate}
  \item \textsc{BFKLex}~\cite{Chachamis:2011nz,Chachamis:2012fk,Chachamis:2015ico,deLeon:2020myv,Baldenegro:2024ndr}, a  BFKL--based Monte Carlo that generates gluon ladders ordered in rapidity but not in transverse momentum, convoluted with proton PDFs and clustered into jets.
  \item \textsc{PYTHIA8}~\cite{Bierlich:2022pfr}, representing a standard DGLAP parton shower with $p_T$ ordering, multiple parton interactions (MPI) and Lund string hadronization.
  \item \textsc{HERWIG7}~\cite{Bahr:2008pv,Bellm:2015jjp}, with an angular--ordered DGLAP shower, MPI and cluster hadronization.
\end{enumerate}
Unless otherwise specified, default tunes and PDF sets are used.
The \textsc{BFKLex} events are generated at leading
logarithmic (LL) accuracy. For the DGLAP side we  compare to 
LL parton showers as implemented in \textsc{PYTHIA8} and \textsc{HERWIG7}. 
Including fixed-order NLO matrix elements for low jet multiplicities 
(e.g.\ via \textsc{POWHEG} dijet samples matched to showers) would improve 
the description of the first few jets but would treat higher multiplicities 
inhomogeneously, which is undesirable in an analysis that explicitly 
mixes events from all jet multiplicities. For our purposes, a comparison between LL BFKL 
evolution and standard LL DGLAP showers is therefore the most transparent approach.
Since we investigate the structure of radiation across the entire rapidity interval, 
rather than just the hardest emissions, the distinct ordering criteria of the evolution equations 
(BFKL diffusion vs.\ DGLAP collinear ordering) are the dominant shaping mechanisms, 
rendering fixed--order corrections subleading for these shape observables.

We have verified that turning hadronization on and off in \textsc{PYTHIA8} and \textsc{HERWIG7} 
has a negligible impact on the \emph{shape--normalized} rank distributions. 
We also find that, in all three generators, the rank shapes are stable under variations of the 
jet radius $R\in\{0.4,0.5,0.6\}$ and under replacing the default proton PDF with alternative sets.
These variations affect overall rates 
but leave the rank shapes unchanged within the Monte Carlo statistical precision. We further note that raising the 
transverse momentum threshold to $p_{T,\min}=30$~GeV---a standard requirement to 
mitigate pileup in high--luminosity LHC environments---does not degrade the discrimination power 
of these observables, as the topological differences are driven by the rapidity 
ordering rather than the softest scale.

\section{Results}

\begin{figure}[t]
  \centering
  \begin{subfigure}[t]{0.48\linewidth}
    \centering
    \includegraphics[width=\linewidth]{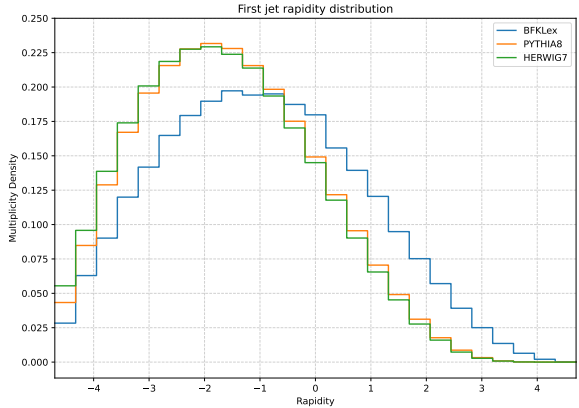}
    \caption{Jet1 (most backward)}
  \end{subfigure}\hfill
  \begin{subfigure}[t]{0.48\linewidth}
    \centering
    \includegraphics[width=\linewidth]{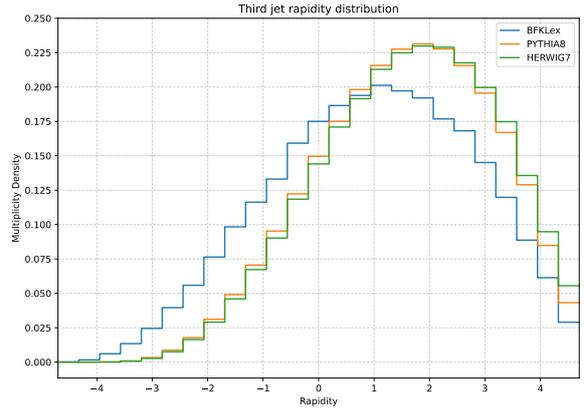}
    \caption{Jet3 (most forward)}
  \end{subfigure}

  \vspace{0.8em}

  \begin{subfigure}[t]{0.48\linewidth}
    \centering
    \includegraphics[width=\linewidth]{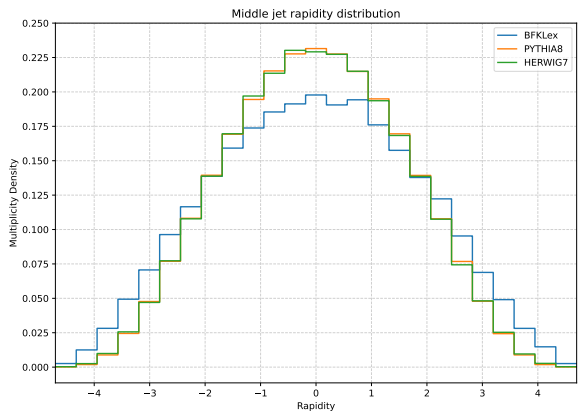}
    \caption{Jet2 (central)}
  \end{subfigure}
  \caption{\textbf{8~TeV, fixed multiplicity $N{=}3$.} Normalized to unit area rapidity distributions for Jet1, Jet2 and Jet3 (ordered by rapidity) from \textsc{BFKLex}, \textsc{PYTHIA8} and \textsc{HERWIG7}. For all three ranks  we see a clear separation: \textsc{BFKLex} fills the inter--tag interval more democratically, while the DGLAP showers exhibit comparatively stronger edge enhancement (Jet1, Jet3) and narrower central distribution (Jet2).}
  \label{fig:N3-8TeV}
\end{figure}

Figure~\ref{fig:N3-8TeV} shows the 8~TeV results for fixed multiplicity $N=3$. 
All three ranked rapidity distributions differ clearly between \textsc{BFKLex} 
and the DGLAP showers. For Jet1 and Jet3, \textsc{BFKLex} populates the backward 
and forward edges more democratically, while \textsc{PYTHIA8} and \textsc{HERWIG7} 
show comparatively stronger edge enhancement. Jet2 also exhibits a marked 
separation: \textsc{BFKLex} yields a broader, more plateau--like central distribution, 
whereas the DGLAP showers push the central jet in a narrower zone in the middle.

\begin{figure}[t]
  \centering
  \begin{subfigure}{0.48\linewidth}
    \includegraphics[width=\linewidth]{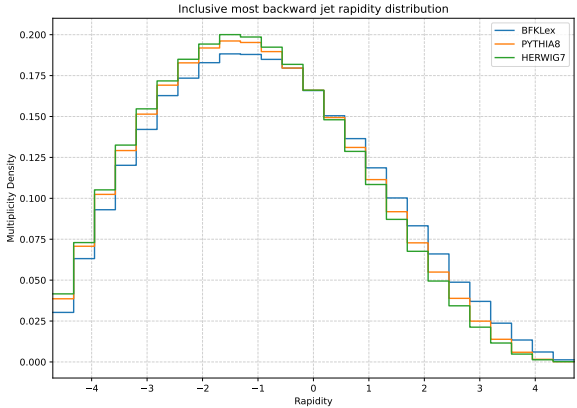}
    \caption{MB (most backward)}
  \end{subfigure}\hfill
  \begin{subfigure}{0.48\linewidth}
    \includegraphics[width=\linewidth]{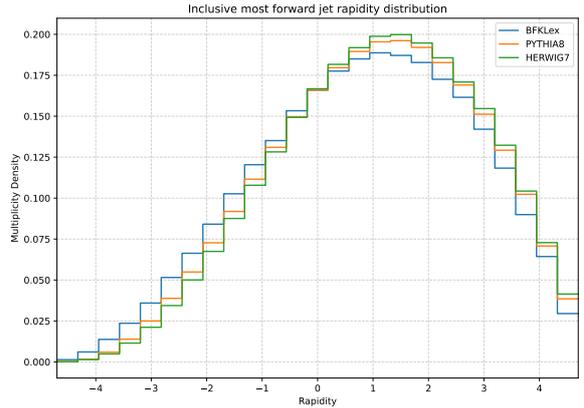}
    \caption{MF (most forward)}
  \end{subfigure}
  \caption{\textbf{8~TeV, across multiplicities.} Normalized rapidity distributions for the most--backward (MB) and most--forward (MF) jets. The three generators give very similar shapes, dominated by PDFs and acceptance.}
  \label{fig:MBMF-8TeV}
\end{figure}

\begin{figure}[t]
  \centering
  \begin{subfigure}{0.48\linewidth}
    \includegraphics[width=\linewidth]{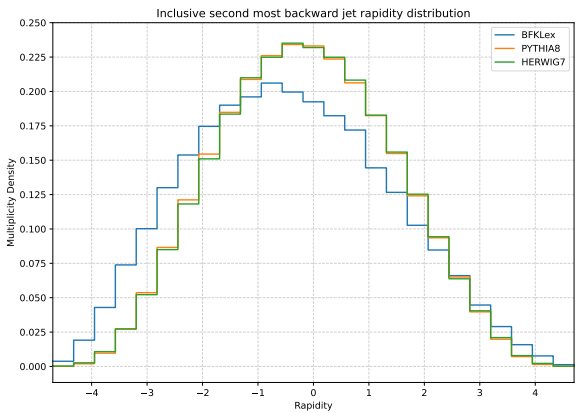}
    \caption{SMB (second--most backward)}
  \end{subfigure}\hfill
  \begin{subfigure}{0.48\linewidth}
    \includegraphics[width=\linewidth]{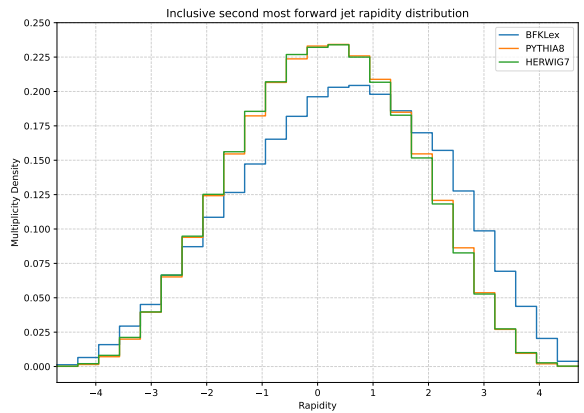}
    \caption{SMF (second--most forward)}
  \end{subfigure}
  \caption{\textbf{8~TeV, across multiplicities.} Normalized rapidity distributions for the second--most--backward (SMB) and second--most--forward (SMF) jets. These inner ranks show the largest separation between \textsc{BFKLex} and the DGLAP showers: \textsc{PYTHIA8} and \textsc{HERWIG7} exhibit more central weight, while \textsc{BFKLex} remains broader with a much more pronounced left (backward jet) and right (forward jet) skewness.}
  \label{fig:SMFSMB-8TeV}
\end{figure}

Figures~\ref{fig:MBMF-8TeV} and~\ref{fig:SMFSMB-8TeV} 
display the across--multiplicity results at 8~TeV. The MB and MF distributions
 are again very similar between \textsc{BFKLex}, \textsc{PYTHIA8} and \textsc{HERWIG7}, 
 as expected for the extremal ranks that are most constrained by the tagging and acceptance. 
 In contrast, the SMB and SMF distributions show the clearest separation: \textsc{BFKLex} 
 produces broader distributions, with clear left and right skewness whereas the DGLAP showers 
 yield distributions with more pronounced central weight.

\begin{figure}[t]
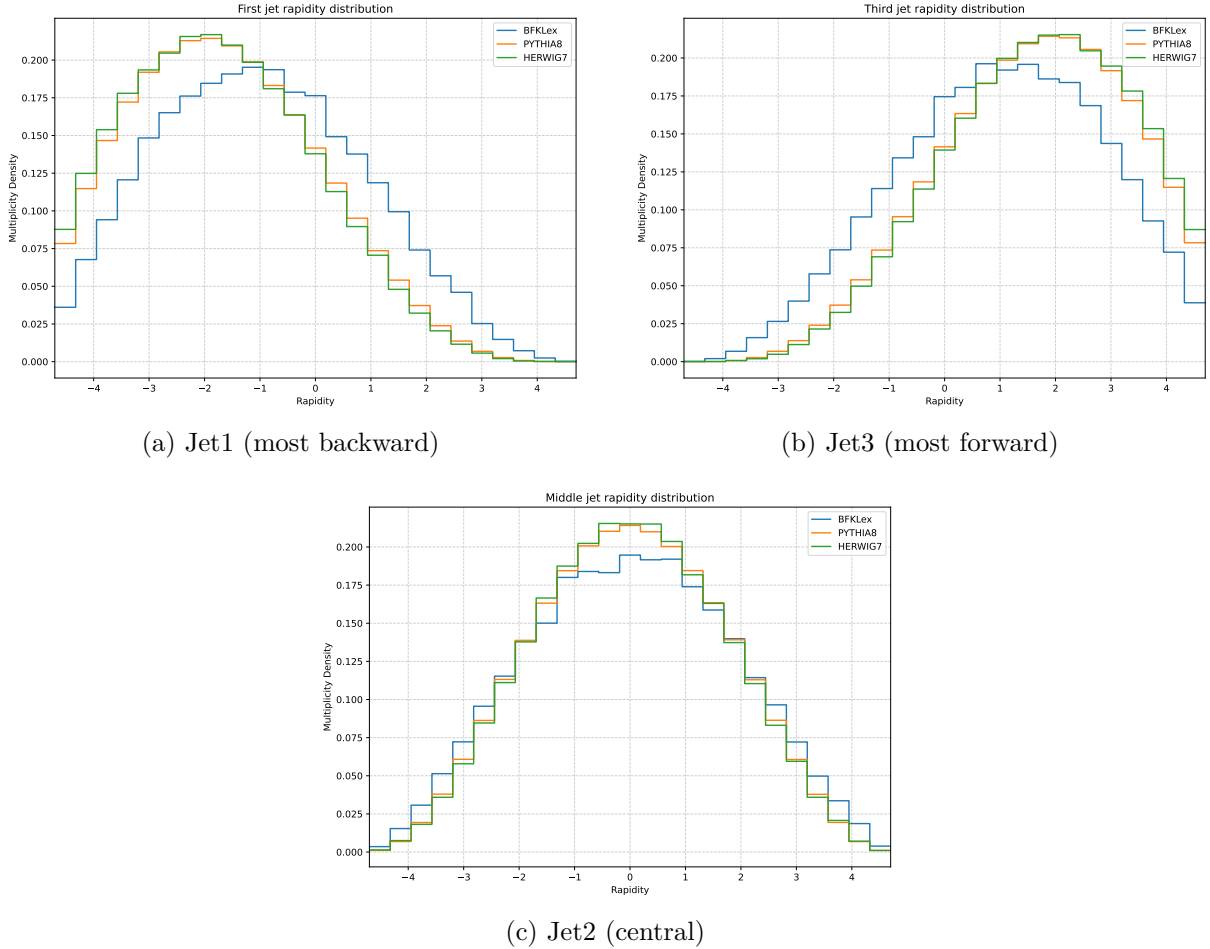

  \centering
  \begin{subfigure}[t]{0.48\linewidth}
    \centering
    \includegraphics[width=\linewidth]{\thirteentev{rapidity_distribution_jet1.pdf}}
    \caption{Jet1 (most backward)}
  \end{subfigure}\hfill
  \begin{subfigure}[t]{0.48\linewidth}
    \centering
    \includegraphics[width=\linewidth]{\thirteentev{rapidity_distribution_jet3.pdf}}
    \caption{Jet3 (most forward)}
  \end{subfigure}

  \vspace{0.8em}

  \begin{subfigure}[t]{0.48\linewidth}
    \centering
    \includegraphics[width=\linewidth]{\thirteentev{rapidity_distribution_jet2.pdf}}
    \caption{Jet2 (central)}
  \end{subfigure}
  \caption{\textbf{13~TeV, fixed multiplicity $N{=}3$.} Normalized rapidity distributions for Jet1, Jet2 and Jet3. The central rank again shows a clear separation between \textsc{BFKLex} and the DGLAP showers.}
  \label{fig:N3-13TeV}
\end{figure}

At $\sqrt{s}=13$~TeV we find the same qualitative pattern. 
Figure~\ref{fig:N3-13TeV} shows the $N{=}3$ distributions: all three ranks 
display clear shape differences between \textsc{BFKLex} and the DGLAP showers. 
As at 8~TeV, \textsc{BFKLex} fills the inter--tag rapidity interval more 
democratically, while \textsc{PYTHIA8} and \textsc{HERWIG7} exhibit stronger 
edge enhancement for the outer jets and a narrower central distribution for Jet2.

\begin{figure}[t]
  \centering
  \begin{subfigure}{0.48\linewidth}
    \includegraphics[width=\linewidth]{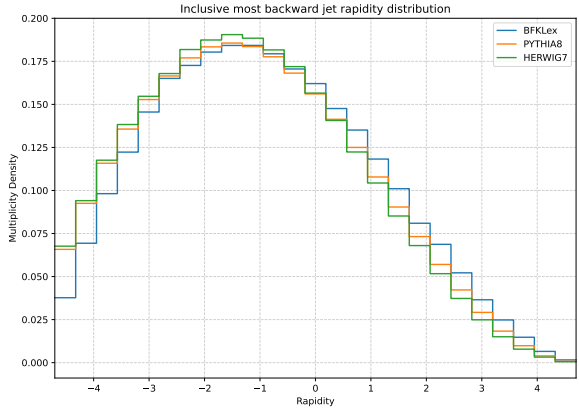}
    \caption{MB (most backward)}
  \end{subfigure}\hfill
  \begin{subfigure}{0.48\linewidth}
    \includegraphics[width=\linewidth]{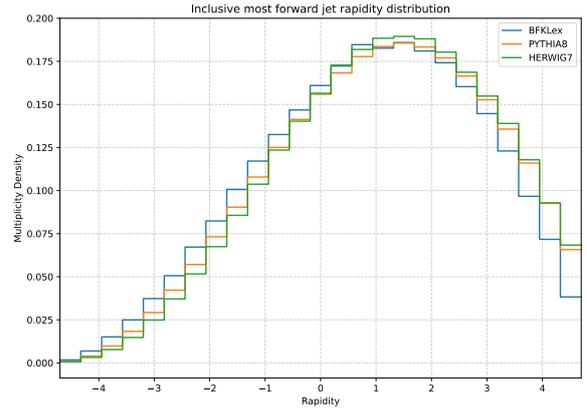}
    \caption{MF (most forward)}
  \end{subfigure}
  \caption{\textbf{13~TeV, across multiplicities.} Normalized rapidity distributions for the MB and MF jets. As at 8~TeV, the three generators give very similar shapes.}
  \label{fig:MBMF-13TeV}
\end{figure}

\begin{figure}[t]
  \centering
  \begin{subfigure}{0.48\linewidth}
    \includegraphics[width=\linewidth]{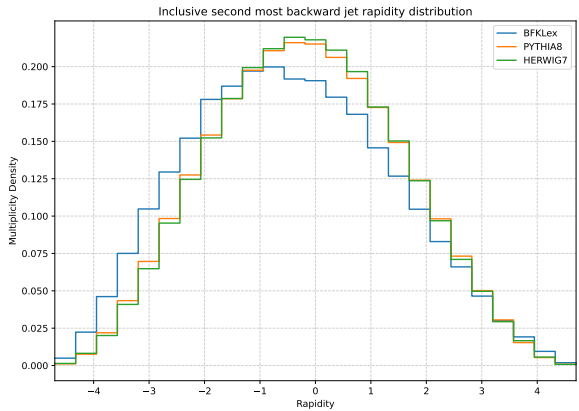}
    \caption{SMB (second--most backward)}
  \end{subfigure}\hfill
  \begin{subfigure}{0.48\linewidth}
    \includegraphics[width=\linewidth]{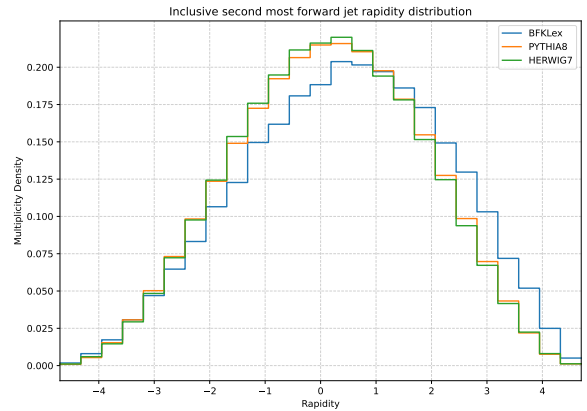}
    \caption{SMF (second--most forward)}
  \end{subfigure}
  \caption{\textbf{13~TeV, across multiplicities.} Normalized rapidity distributions for SMB and SMF jets. Inner ranks remain the most sensitive probes of the underlying rapidity law. The separation between \textsc{BFKLex} and the DGLAP showers persists at 13~TeV.}
  \label{fig:SMFSMB-13TeV}
\end{figure}

We have also repeated the analysis in bins of the rapidity span $\Delta Y$ 
between the two extremal jets, imposing minimum separations $\Delta Y \ge 1,3,5$. 
The MB/MF distributions move closer to the edges by construction, 
while the SMB/SMF shapes retain their discriminating power between \textsc{BFKLex} 
and the DGLAP showers in all $\Delta Y$ bins where statistics remain healthy. 
This supports the view that rank--ordered observables provide 
complementary information to standard Mueller--Navelet angular harmonics in the same event samples.

\section{Discussion and outlook}

Rank--ordered rapidity distributions are standard objects in 
mathematical statistics (order statistics) but have not, to our 
knowledge, been used as primary observables in collider QCD. 
Here we have shown that, when applied to multijet final states 
in inclusive dijet topologies, they provide simple, experimentally 
robust and theoretically interpretable probes of the underlying rapidity production law.

Within an approximate i.i.d.\ picture, the mapping in Eq.~\eqref{eq:order} 
guarantees that the shapes of the rank distributions are direct images of the 
effective parent density realized by a given dynamics. In practice, 
effects such as global recoil, color coherence, transverse--momentum 
thresholds and acceptance boundaries simply deform this effective parent. 
Our Monte Carlo study suggests that rank shapes are largely insensitive to 
nonperturbative details (MPI, hadronization) and to standard variations of 
jet radius and PDFs, but respond strongly to differences between BFKL--like 
and DGLAP--like radiation patterns.

A natural next step would be to extend the present analysis to 
next-to-leading logarithmic (NLL) BFKL accuracy, including NLO jet vertices. 
This is technically demanding, and lies beyond the scope of this first study. 

Preliminary checks at the level of the Green's function in related work 
suggest that going from LL to NLL mainly reshuffles the overall 
normalizations and energy dependence and does not qualitatively alter 
the pattern by which high--energy radiation fills rapidity space.
Furthemore, because the rank--ordered observables proposed here are 
shape--normalized to unit area, they are immune to the large $K$--factors 
that typically differentiate LL from NLL cross sections. They are instead 
sensitive to the diffusive properties of the evolution kernel, which persist at NLL.
While we cannot yet make a similarly firm statement about the full 
NLO impact factors, we do not expect the rank--ordered rapidity patterns 
found here to disappear at NLL: they are a direct imprint of the underlying 
high--energy radiation mechanism rather than a peculiarity of a specific 
logarithmic truncation. A detailed NLL study is therefore well motivated, 
but is a substantial project in its own right.

From a broader perspective, the main message of this work is conceptual as 
much as it is technical. Interpreting the rapidity distributions
as order statistics of an effective parent rapidity density provides a remarkably simple way to 
expose differences in multi--jet radiation patterns that are invisible in more 
conventional one--dimensional observables.
In our Monte Carlo study, a BFKL--based generator 
and DGLAP showers produce very similar shapes for the outer ranks (MB, MF) across multiplicities, 
yet differ clearly in two ways: (i) in the distributions of the more central ranks (SMB, SMF) 
when summing over all jet multiplicities, and (ii) in all ranks for fixed jet multiplicity ($N{=}3$). 
This is precisely the kind of ``hidden'' information that rank--ordered observables are designed to retain.

Because these observables are just normalized one--dimensional histograms 
of standard jets, they can be measured in existing inclusive dijet samples, 
including but not limited to classic Mueller--Navelet selections with large 
rapidity separation between the two extremal jets. In particular, they can 
reuse the same triggers and event samples as current azimuthal--decorrelation 
and dijet measurements at the LHC. We therefore 
view the present work as a proof of concept for a wider program: using 
order statistics as a systematic language to turn multiperipheral and BFKL 
intuition about ``how rapidity space is filled'' into concrete, data--driven tests. 
It will be particularly interesting to confront the predictions shown here with Run~2 and Run~3 
measurements, to refine the comparison with alternative high--energy 
frameworks, and to explore extensions such as joint distributions of multiple ranks and 
lower $p_T$ thresholds with track--based jets.

\medskip
\noindent\textbf{Acknowledgments.} \,
The work of G.~Chachamis was supported by the Fundação para a Ciência e a Tecnologia (FCT, Portugal) through project EXPL/FIS-PAR/1195/2021 (\url{http://doi.org/10.54499/EXPL/FIS-PAR/1195/2021}), project CERN/FIS-PAR/0032/2021 (\url{http://doi.org/10.54499/CERN/FIS-PAR/0032/2021}), and under the ``Investigador FCT'' contract 03216/2017. The work of D.~Vaccaro was supported by the FCT through project EXPL/FIS-PAR/1195/2021 and by the PT-CERN PhD Grant (Call 2021/2).

A.~Sabio Vera acknowledges support from the Spanish Agencia Estatal de Investigación, grants PID2022-142545NB-C22 and IFT Centro de Excelencia Severo Ochoa CEX2020-001007-S, funded by MCIN/AEI/10.13039/501100011033 and by ERDF A way of making Europe; and from the European Union's Horizon 2020 research and innovation programme under grant agreement No.~824093.

\bibliographystyle{ieeetr}
\bibliography{refs}

@article{Chachamis:2015ico,
    author = "Chachamis, G. and Sabio Vera, A.",
    title = "{The high-energy radiation pattern from BFKLex with double-log collinear contributions}",
    eprint = "1512.03603",
    archivePrefix = "arXiv",
    primaryClass = "hep-ph",
    doi = "10.1007/JHEP02(2016)064",
    journal = "JHEP",
    volume = "02",
    pages = "064",
    year = "2016"
}

@article{Baldenegro:2024ndr,
    author = "Baldenegro, C. and Chachamis, G. and Kampshoff, M. and Klasen, M. and Milhano, G. J. and Royon, C. and Sabio Vera, A.",
    title = "{Multijet event shape variables for Mueller-Navelet jet topologies}",
    eprint = "2406.10681",
    archivePrefix = "arXiv",
    primaryClass = "hep-ph",
    doi = "10.1103/PhysRevD.110.114027",
    journal = "Phys. Rev. D",
    volume = "110",
    number = "11",
    pages = "114027",
    year = "2024"
}

@article{Gribov:1972ri,
    author = "Gribov, V. N. and Lipatov, L. N.",
    title = "{Deep inelastic e p scattering in perturbation theory}",
    reportNumber = "IPTI-381-71",
    journal = "Sov. J. Nucl. Phys.",
    volume = "15",
    pages = "438--450",
    year = "1972"
}

@article{Gribov:1972rt,
    author = "Gribov, V. N. and Lipatov, L. N.",
    title = "{e+ e- pair annihilation and deep inelastic e p scattering in perturbation theory}",
    journal = "Sov. J. Nucl. Phys.",
    volume = "15",
    pages = "675--684",
    year = "1972"
}

@article{Lipatov:1974qm,
    author = "Lipatov, L. N.",
    title = "{The parton model and perturbation theory}",
    journal = "Yad. Fiz.",
    volume = "20",
    pages = "181--198",
    year = "1974"
}

@article{Altarelli:1977zs,
    author = "Altarelli, Guido and Parisi, G.",
    title = "{Asymptotic Freedom in Parton Language}",
    reportNumber = "LPTENS-77-6",
    doi = "10.1016/0550-3213(77)90384-4",
    journal = "Nucl. Phys. B",
    volume = "126",
    pages = "298--318",
    year = "1977"
}

@article{Dokshitzer:1977sg,
    author = "Dokshitzer, Yuri L.",
    title = "{Calculation of the Structure Functions for Deep Inelastic Scattering and e+ e- Annihilation by Perturbation Theory in Quantum Chromodynamics.}",
    journal = "Sov. Phys. JETP",
    volume = "46",
    pages = "641--653",
    year = "1977"
}

@article{Lipatov:1985uk,
    author = "Lipatov, L. N.",
    title = "{The Bare Pomeron in Quantum Chromodynamics}",
    reportNumber = "LENINGRAD-85-1137",
    journal = "Sov. Phys. JETP",
    volume = "63",
    pages = "904--912",
    year = "1986"
}

@article{Balitsky:1978ic,
    author = "Balitsky, I. I. and Lipatov, L. N.",
    title = "{The Pomeranchuk Singularity in Quantum Chromodynamics}",
    journal = "Sov. J. Nucl. Phys.",
    volume = "28",
    pages = "822--829",
    year = "1978"
}

@article{Kuraev:1977fs,
    author = "Kuraev, E. A. and Lipatov, L. N. and Fadin, Victor S.",
    title = "{The Pomeranchuk Singularity in Nonabelian Gauge Theories}",
    journal = "Sov. Phys. JETP",
    volume = "45",
    pages = "199--204",
    year = "1977"
}

@article{Kuraev:1976ge,
    author = "Kuraev, E. A. and Lipatov, L. N. and Fadin, Victor S.",
    title = "{Multi - Reggeon Processes in the Yang-Mills Theory}",
    journal = "Sov. Phys. JETP",
    volume = "44",
    pages = "443--450",
    year = "1976"
}

@article{Lipatov:1976zz,
    author = "Lipatov, L. N.",
    title = "{Reggeization of the Vector Meson and the Vacuum Singularity in Nonabelian Gauge Theories}",
    journal = "Sov. J. Nucl. Phys.",
    volume = "23",
    pages = "338--345",
    year = "1976"
}

@article{Fadin:1975cb,
    author = "Fadin, Victor S. and Kuraev, E. A. and Lipatov, L. N.",
    title = "{On the Pomeranchuk Singularity in Asymptotically Free Theories}",
    doi = "10.1016/0370-2693(75)90524-9",
    journal = "Phys. Lett. B",
    volume = "60",
    pages = "50--52",
    year = "1975"
}

@article{deLeon:2020myv,
    author = "de Le\'on, N. Bethencourt and Chachamis, G. and Sabio Vera, A.",
    title = "{Multiperipheral final states in crowded twin-jet events at the LHC}",
    eprint = "2012.09664",
    archivePrefix = "arXiv",
    primaryClass = "hep-ph",
    doi = "10.1016/j.nuclphysb.2021.115518",
    journal = "Nucl. Phys. B",
    volume = "971",
    pages = "115518",
    year = "2021"
}

@article{Chachamis:2012fk,
    author = "Chachamis, G. and Sabio Vera, A.",
    title = "{The NLO N =4 SUSY BFKL Green function in the adjoint representation}",
    eprint = "1206.3140",
    archivePrefix = "arXiv",
    primaryClass = "hep-th",
    doi = "10.1016/j.physletb.2012.09.051",
    journal = "Phys. Lett. B",
    volume = "717",
    pages = "458--461",
    year = "2012"
}

@article{Chachamis:2011nz,
    author = "Chachamis, G. and Sabio Vera, A.",
    title = "{The Colour Octet Representation of the Non-Forward BFKL Green Function}",
    eprint = "1112.4162",
    archivePrefix = "arXiv",
    primaryClass = "hep-th",
    doi = "10.1016/j.physletb.2012.02.036",
    journal = "Phys. Lett. B",
    volume = "709",
    pages = "301--308",
    year = "2012"
}

@article{Mueller:1986ey,
    author = "Mueller, Alfred H. and Navelet, H.",
    title = "{An Inclusive Minijet Cross-Section and the Bare Pomeron in QCD}",
    reportNumber = "SACLAY-SPHT-86-094",
    doi = "10.1016/0550-3213(87)90705-X",
    journal = "Nucl. Phys. B",
    volume = "282",
    pages = "727--744",
    year = "1987"
}

@article{Cacciari:2011ma,
    author = "Cacciari, Matteo and Salam, Gavin P. and Soyez, Gregory",
    title = "{FastJet User Manual}",
    eprint = "1111.6097",
    archivePrefix = "arXiv",
    primaryClass = "hep-ph",
    reportNumber = "CERN-PH-TH-2011-297",
    doi = "10.1140/epjc/s10052-012-1896-2",
    journal = "Eur. Phys. J. C",
    volume = "72",
    pages = "1896",
    year = "2012"
}

@article{Bierlich:2022pfr,
    author = "Bierlich, Christian and others",
    title = "{A comprehensive guide to the physics and usage of PYTHIA 8.3}",
    eprint = "2203.11601",
    archivePrefix = "arXiv",
    primaryClass = "hep-ph",
    reportNumber = "LU-TP 22-16, MCNET-22-04, FERMILAB-PUB-22-227-SCD",
    doi = "10.21468/SciPostPhysCodeb.8",
    journal = "SciPost Phys. Codeb.",
    volume = "2022",
    pages = "8",
    year = "2022"
}

@article{Bahr:2008pv,
    author = "Bahr, M. and others",
    title = "{Herwig++ Physics and Manual}",
    eprint = "0803.0883",
    archivePrefix = "arXiv",
    primaryClass = "hep-ph",
    reportNumber = "CERN-PH-TH-2008-038, CAVENDISH-HEP-08-03, KA-TP-05-2008, DCPT-08-22, IPPP-08-11, CP3-08-05",
    doi = "10.1140/epjc/s10052-008-0798-9",
    journal = "Eur. Phys. J. C",
    volume = "58",
    pages = "639--707",
    year = "2008"
}

@article{Bellm:2015jjp,
    author = "Bellm, Johannes and others",
    title = "{Herwig 7.0/Herwig++ 3.0 release note}",
    eprint = "1512.01178",
    archivePrefix = "arXiv",
    primaryClass = "hep-ph",
    reportNumber = "CERN-PH-TH-2015-289, MAN-HEP-2015-15, IFJPAN-IV-2015-13, KA-TP-18-2015, DCPT-15-142, MCNET-15-28, IPPP-15-71, HERWIG-2015-01",
    doi = "10.1140/epjc/s10052-016-4018-8",
    journal = "Eur. Phys. J. C",
    volume = "76",
    number = "4",
    pages = "196",
    year = "2016"
}

\end{document}